\documentstyle[twocolumn]{jpsj}

\def\runtitle{Electronic Processes 
at the Breakdown of the Quantum Hall Effect}
\def\runauthor{Hiroshi {\sc Akera}}

\title{Electronic Processes at the Breakdown of the Quantum Hall Effect
\footnote{submitted to the Journal of the Physical Society of Japan}}

\author{Hiroshi {\sc Akera}\footnote{E-mail: akera@eng.hokudai.ac.jp}}

\inst{Department of Applied Physics, Hokkaido University, 
Sapporo 060-8628}

\recdate{June 29, 2000}

\abst
{Microscopic processes giving the energy gain and loss of 
a two-dimensional electron system in long-range potential fluctuations  
are studied theoretically 
at the breakdown of the quantum Hall effect 
in the case of even-integer filling factors.  
The Coulomb scattering within a broadened Landau level is proposed  
to give the gain, 
while the phonon scattering to give the loss. 
The energy balance equation shows that 
the electron temperature $T_e$ and the diagonal conductivity $\sigma_{xx}$
exhibit a bistability above the lower critical electric field $E_{c1}$.  
Calculated values of $E_{c1}$ as well as  
$T_e$ and $\sigma_{xx}$ at $E_{c1}$ are 
in agreement with the observed values in their orders of magnitude. 
}

\kword{integer quantum Hall effect, nonlinear transport, theory}
\begin{document}
\sloppy
\maketitle

\def\vecj{\mib j}
\def\vecE{\mib E}
\def\vecr{\mib r}
\def\vecq{\mib q}
\def\ve{\varepsilon}
\def\lvh{l_{\rm vh}}
\def\vt{\tilde v}
%
%++++++++++++++++++++++++++++++++++++++++++++++++++++
In the quantum Hall effect~\cite{Klitzing80,Kawaji81} (QHE), 
the diagonal conductivity $\sigma_{xx}$
is vanishingly small in the low-current regime. 
When the current is increased up to a critical value, 
$\sigma_{xx}$ increases by several orders of magnitude 
within a narrow range of the current 
and the QHE breaks down.~\cite{Ebert83,Cage83,Kuchar84}  
In spite of extensive studies on the breakdown of QHE,~\cite{review} 
the mechanism of the breakdown has not been fully understood.  
In many samples, 
$\sigma_{xx}$ increases discontinuously at the critical current 
and exhibits a hysteresis as a function of the current,   
which suggests that the breakdown of QHE belongs to  
the nonequilibrium phase transition. 
The phases below and above the transition can be considered to be 
homogeneous in a large class of samples, 
since the critical current is proportional to 
the sample width.~\cite{Kawaji93,Boisen94} 
In this paper, we study theoretically 
the nonequilibrium phase transition between homogeneous states
at the breakdown of the quantum Hall effect 
in the case of even-integer filling factors.  

Among a variety of theories~\cite{Ebert83,Trugman83,Gurevich84,Heinonen84,
Streda84,Stormer85,Smrcka85,Komiyama85,Eaves86,Riess93,Tsemekhman97,
Chaubet98,Eaves98,Ishikawa98,Shizuya99,review} 
proposed for the mechanism of the breakdown of QHE, 
the hot-electron theory~\cite{Ebert83,Gurevich84,Komiyama85} 
showed the existence of 
the hysteresis~\cite{Gurevich84,Komiyama85} and 
reproduced the observed abrupt change of $\sigma_{xx}$ 
with the current density.~\cite{Komiyama85} 
However, this theory~\cite{Komiyama85} employed an observed result 
to obtain the energy gain 
$\vecj \!\cdot\! \vecE=\sigma_{xx} E^2$ 
($\vecj$: current density, $\vecE$: electric field) 
and did not study the microscopic process of $\sigma_{xx}$.  
Studies~\cite{Heinonen84,Chaubet98} 
were made for the process of $\sigma_{xx}$ 
by considering an inter-Landau-level phonon scattering. 
The calculated $\sigma_{xx}$, however, 
had no hysteresis as a function of the electric field and 
the calculated critical electric field 
was at least one order of magnitude larger than the observed one. 
The microscopic process of the energy gain and  
the energy dissipation remains unsolved.  

In this paper we propose a Coulomb scattering 
within a Landau level in a slowly-varying potential
as the dominant electronic process 
giving the energy gain. 
The potential in the plane of the two-dimensional electron system (2DES)
is fluctuating due to ionized donors in the layer several hundred \AA\  above 
2DES, and its importance in the breakdown of QHE has been discussed 
in the literature.~\cite{Trugman83,Streda84,Smrcka85,Tsemekhman97} 
Fluctuations are in a scale of $\lvh=0.1{\rm \mu m}$ 
($\lvh$ is the distance between a potential hill and a neighboring valley)
and have a standard deviation of $20 {\rm meV}$ 
before the screening.~\cite{Nixon90}
The screening does not completely wash out the fluctuations 
in strong magnetic fields due to the discrete nature of the energy spectrum. 
The screened potential has a reduced width equal to 
the Landau level separation~\cite{Wulf88,Efros88} 
and, even in a filling factor of $2N$($N=1,2,\cdots$), 
electrons (holes) populate in the $N+1$th ($N$th) Landau level 
(we assume the spin degeneracy). 
Therefore the energy dissipation occurs within a Landau level 
and it is much larger than that due to inter-Landau-level scatterings.
In such a slowly-varying potential in strong magnetic fields, 
closed orbits are formed around its hills and valleys,  
and their typical size is of the order of $\lvh$ and much larger than 
the magnetic length $l$ since $\lvh \sim 10 l$ in $B=5{\rm T}$.
A hill orbit at the center of the Landau level is 
in close proximity to neighboring valley orbits 
and hoppings between these orbits along the electric field are 
the most dominant process of the energy gain. 
We propose a Coulomb scattering as the dominant mechanism of a hopping, 
in which an electron hops from a hill (valley) orbit to a valley (hill) orbit 
and, at the same time, one of other electrons is excited (relaxed). 
The dominant process for the energy loss, on the other hand, is given by 
acoustic-phonon deformation-potential scatterings, again, 
within the Landau level. 
In our calculations, 
(1) we consider only the activation transport at the Landau-level center 
and neglect the tunneling transport at the Landau-level edge, 
which limits the quantitative validity of our calculations to 
the higher $T_e$ branch with the larger $\sigma_{xx}$.
We also assume 
(2) the vanishing lattice temperature: $T_L=0$, and 
(3) even-integer filling factors.  
Our calculations show 
that the electron temperature $T_e$ and $\sigma_{xx}$ exhibit a bistability 
above the lower critical electric field $E_{c1}$, 
giving a hysteresis as a function of $E$.  
Calculated orders of magnitude of 
$E_{c1}$ as well as $T_e$ and $\sigma_{xx}$ at $E_{c1}$ 
agree with the experimental ones. 

%++++++++++++++++++++++++++++++++++++++++++++++++++++
In a steady state, the energy balance equation 
%====================================================
\begin{equation}
P_G(T_e,E)=P_L(T_e,T_L) ,
\end{equation}
%====================================================
holds for the energy gain $P_G$ and the loss $P_L$, 
and $P_G$ is related with the diagonal conductivity $\sigma_{xx}$ by 
%====================================================
\begin{equation}
P_G(T_e,E)=\vecj \!\cdot\! \vecE=\sigma_{xx} (T_e) E^2 .
\end{equation}
%====================================================
We assume, 
following the previous works,~\cite{Ebert83,Gurevich84,Komiyama85}
that $\sigma_{xx}$ depends on $E$ only through $T_e(E)$, 
which is consistent with the experiment.~\cite{Komiyama96}
The electron temperature is obtained as a function of $E$ 
from the energy balance equation. 

The electron distribution function $f(\ve)$ is given by
%====================================================
\begin{equation}
f(\ve)= {1 \over \exp[(\ve-\mu)/k_B T_e] +1} .
\end{equation}
%====================================================
Since the filling factor is an even integer, $2N$, 
and the spin splitting is much smaller than $k_B T_e$, 
the chemical potential $\mu$ is at the middle point between 
the $N$th and the $N \!+\! 1$th Landau levels, and
the electron distribution in the $N \!+\! 1$th Landau level and the hole 
distribution in the $N$th Landau level for both spins 
are described by the same function. 
In the following the zero of energy is taken at the center of 
the $N \!+\! 1$th Landau level and $\mu=-\hbar\omega_c/2$ 
with $\omega_c$ the cyclotron frequency. 
The electron (hole) occupation of current-carrying states around 
the $N \!+\! 2$th ($N-1$th) Landau-level center, $f(\hbar\omega_c)$, 
is neglected, 
which means that we restrict our calculations to  
a lower-$T_e$ range of the higher-$T_e$ branch. 

%#############################################################
\begin{figure}
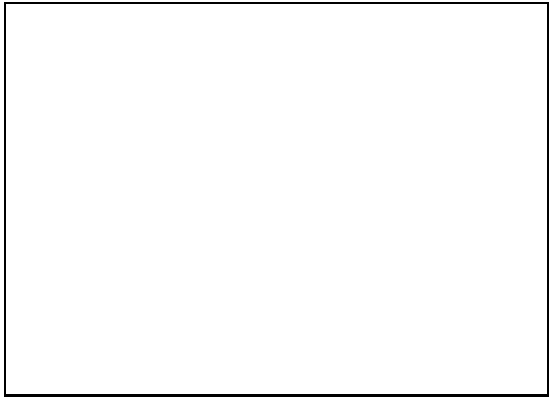

\figureheight{5cm}
\caption{
A Coulomb scattering among 
one hill orbit $\ve_3$ and three valley orbits $\ve_1, \ve_2, \ve_4$
in a slowly-varying potential $V$.  
$\Delta \mu $ is the chemical potential difference 
between the hill and the valley.
} 
\label{fig:potential}
\end{figure}
%#############################################################

%++++++++++++++++++++++++++++++++++++++++++++++++++++
The proposed electronic process for the energy gain 
is a Coulomb scattering, as illustrated in Fig.~\ref{fig:potential}, 
from a hill orbit $\ve_3$ to a valley orbit $\ve_4$ 
with an excitation from $\ve_1$ to $\ve_2$ 
(we use the energy to label an orbit 
since we consider processes within one hill and one valley). 
We consider the excitation at the closest valley and hill from 
the hopping electron. 
Excitations at larger distances are negligible 
since the Coulomb matrix element, 
which is approximately the interaction between a point charge and a 
dipole, 
decreases with the distance $d$ as $d^{-2}$ 
and the transition rate as $d^{-4}$. 
The guiding-center distance $\Delta X$ corresponding to 
the energy difference $\Delta \ve = \ve_2 \!-\! \ve_1 = \ve_3 \!-\! \ve_4$ 
is 
$\Delta X=\Delta \ve / eE_l$
with $E_l$ the local electric field strength due to the slowly-varying potential. 
The transition rate 
is appreciable only when $\Delta X \sim l$ or smaller,  
since the wavefunction overlap integral 
$S(\Delta X)$ (along the perpendicular direction) decreases with $\Delta X$ as 
%====================================================
\begin{equation}
S(\Delta X)= \exp(-\Delta X^2/4l^2)=\exp(-\Delta \tilde\ve^2/4 \vt^2) ,
\end{equation}
%====================================================
where $\Delta \tilde\ve=\Delta \ve / \hbar\omega_c$ 
and $\vt=leE_l / \hbar\omega_c$. 
Inter-Landau-level scatterings are neglected here 
since the corresponding $S(\Delta X)$ is much smaller: 
$S(\Delta X)=10^{-11}$ at $\vt=0.1$ 
for a quasi-elastic phonon scattering with $eE_l \Delta X = \hbar\omega_c$ 
and $S(\Delta X)=10^{-3}$ 
for the most dominant Coulomb scattering with $\Delta \ve = \hbar\omega_c /2$
(the transition rate is proportional to $S(\Delta X)^2$ for phonon 
scatterings, and to $S(\Delta X)^4$ for Coulomb scatterings).  
The change of electronic states due to the applied electric field is 
neglected since $E_{c1}/E_l \sim 0.1$. 

The transition rate $W^c_{12,34}$ of the hill-to-valley Coulomb scattering is 
%====================================================
\begin{equation}
W^c_{12,34}=
{2\pi \over \hbar} |\langle \ve_1 \ve_3 \left|  H_c \right| \ve_2 \ve_4 \rangle|^2 
\delta(\ve_1 \!+\! \ve_3 \!-\! \ve_2 \!-\! \ve_4)
\end{equation}
%====================================================
where $H_c$ is the Coulomb interaction:  
$H_c=e^2/\epsilon |\vecr_a - \vecr_b|$ 
with $\epsilon$ the dielectric constant.  
The matrix element of $H_c$ is given by 
%====================================================
\begin{equation}
\left\langle \ve_1 \ve_3 \left|  H_c \right| \ve_2 \ve_4  \right\rangle 
= {e^2 \over \epsilon \lvh} S(\Delta X)^2 f_L I_{\rm orb} .
\end{equation}
%====================================================
$f_L$ is the fraction in length of the orbit $\ve_4$ 
in which the distance to the orbit $\ve_3$ is around $\Delta X$. 
$I_{\rm orb}$ represents the dependence on orbital configurations. 
When the orbits $\ve_1$, $\ve_2$, and $\ve_4$ are 
circles with the common center,  
$I_{\rm orb}$ is a function of the radius $r_1$ of orbit $\ve_1$ as well as 
$\Delta X$ ($r_4=\lvh /2$),
and the square root of the average of $I_{\rm orb}^2$ over $r_1$ and $\Delta X$  
is estimated to be 0.4. 

The energy gain $P_{G}$ per unit area per unit time is 
related by 
%====================================================
\begin{equation}
P_G=  {N_l^2 N_c \over 2 \lvh^2} \langle P_{Gi} \rangle_i
\end{equation}
%====================================================
to the energy gain $P_{Gi}$ 
in the $i$th pair of a hill and a valley 
within the $N\!+\!1$th Landau level with either spin. 
The coordination number $N_c$ is 
the average number of valleys to which an electron hops from a hill. 
$N_l=4$ is the number of possible states for each orbit 
with different spins and Landau indices. 
The energy gain due to a single hill-to-valley hopping is 
$\ve_G=e {\mib E} \!\cdot\! {\Delta \mib r}$ with 
${\Delta \mib r}$ the difference between their position vectors. 
The chemical potential is assumed to be constant within each hill 
and each valley 
and the difference between the hill and the valley to be $\ve_G$.
$P_{Gi}$ is given by 
%====================================================
\begin{eqnarray}
P_{Gi} =&&
\! \sum_{\ve_1,\ve_2,\ve_3,\ve_4} \!\!
f_1 (1\! -\! f_2) \tilde f_3 (1\! -\! f_4)
W^c_{12,34}  \ve_G  \nonumber \\
-&&
\! \sum_{\ve_1,\ve_2,\ve_3,\ve_4} \!\!
(1\! -\! f_1)  f_2 (1\! -\! \tilde f_3)  f_4
W^c_{12,34}  \ve_G
\end{eqnarray}
%====================================================
with $f_n=f(\ve_n)$ $(n=1,2,4)$ and $\tilde f_3 = f(\ve_3-\ve_G)$. 
We make approximations that  
$f(\ve_1) \sim f(\ve_2)$, 
$f(\ve_3-\ve_G) \sim f(\ve_3) -\ve_G f'(\ve_3)$, and $f'(\ve_3) \sim 
f'(0) $, 
by assuming 
$k_B T_e \gg \Delta \ve \sim \vt \hbar\omega_c$ and
$k_B T_e \gg \ve_G$, 
which hold approximately in the higher $T_e$ branch around $E_{c1}$
if we use the value of $T_e(E_{c1})$ and $E_{c1}$ estimated below. 
Then $P_{Gi}$ becomes 
%====================================================
\begin{equation}
P_{Gi}=
\! \sum_{\ve_1,\ve_2,\ve_3,\ve_4} \!\!
f_1 (1\! -\! f_1) [- f'(0)]
W^c_{12,34}  \ve_G^2 .
\end{equation}
%====================================================
In averaging $P_{Gi}$, we change the summation to the integration 
over the energy $\ve$. 
We assume that the density of states $\rho(\ve)$ is slowly varying so that 
$\rho(\ve_1) \sim \rho(\ve_2)$ and 
$\rho(\ve_3) \sim \rho(\ve_4) \sim \rho(0)$. 
With use of $\langle \ve_G^2 \rangle_i = (eE\lvh)^2/2$, 
we obtain 
%====================================================
\begin{equation}
P_G = A_G \tilde E^2 f(0) [1-f(0)] I_f 
\end{equation}
%====================================================
with
%====================================================
\begin{equation}
A_G ={8 \over \pi^2 h } {\lvh^6 \over l^8} 
\left( {e^2 \over \epsilon l} f_L I_{\rm orb} \right)^2 \vt^2 \tilde \rho(0)^2 N_c,
\end{equation}
%====================================================
$\tilde E=eEl / \hbar\omega_c$, and 
%====================================================
\begin{equation}
I_f = - \int^{\infty}_{\mu} d\ve \tilde \rho(\ve)^2 f'(\ve).
\end{equation}
%====================================================
$\rho(\ve)$ is normalized to be unity when integrated within a Landau 
level and $\tilde \rho(\ve)$ is a dimensionless density of states:
$\tilde \rho(\ve) =\rho(\ve) \hbar\omega_c$. 

As the most dominant process of the energy loss at vanishing lattice temperature, 
we consider the acoustic-phonon emission due to the deformation potential.  
The rate $W^p_{12}$ of the phonon emission with a transition 
from $\ve_1$ to $\ve_2$ is 
%====================================================
\begin{equation}
W^p_{12}= {2\pi \over \hbar} \sum_{\vecq}
| C_q |^2 
|\langle \ve_1 \left|  \exp(i \vecq \!\cdot\! \vecr) \right| \ve_2  \rangle|^2 
\delta(\ve_1 - \ve_2 - \hbar\omega_q) 
\end{equation}
%====================================================
where $\vecq=(q_x,q_y,q_z)$ and $\omega_q=c_l q$ are 
the phonon wavevector and angular frequency, respectively, 
with $c_l$ the group velocity of the longitudinal acoustic mode. 
$C_q=qD(\hbar/2 \rho_m V \omega_q)^{1/2}$ with
$D$ the deformation potential, $\rho_m$ the density, and 
$V$ the volume of the sample. 
The matrix element is calculated to give 
%====================================================
\begin{equation}
\int dq_x dq_y 
|\langle \ve_1 \left|  \exp(i \vecq \!\cdot\! \vecr) \right| \ve_2  \rangle|^2 
={(2 \pi)^{3/2} \over L_p  l} {S(\Delta X)^2  \over (1+q_z^2/b^2)^3} .
\end{equation}
%====================================================
with $L_p$ the perimeter of the orbit.
We have used the Fang-Howard wavefunction~\cite{Fang66} 
along $z$ (perpendicular to the plane): 
$\zeta_0(z)=(b^3/2)^{1/2} z \exp(-bz/2)$. 
We here put $q_z=q$ since 
$q/(q_x^2 +q_y^2)^{1/2} \sim l \omega_c \tilde v /c_l \sim 3$ at $B$=5T. 
We also assume that orbits are circles with radius $r_0$ and 
average the transition rate over $r_0$. 

The energy loss $P_{L}$ per unit area per unit time is 
%====================================================
\begin{equation}
P_L=  {N_l \over 2 \lvh^2} \langle P_{Li} \rangle_i
\end{equation}
%====================================================
with
%====================================================
\begin{equation}
P_{Li} =\! \sum_{\ve_1>\ve_2} \!\!
f_1 (1\! -\! f_2) W^p_{12} \cdot (\ve_1-\ve_2) .
\end{equation}
%====================================================
Using the approximations used already in the calculation of $P_G$: 
$f(\ve_1) \sim f(\ve_2)$ and $\rho(\ve_1) \sim \rho(\ve_2)$, 
we obtain 
%====================================================
\begin{equation}
P_L = A_L \tilde T_e I_f
\end{equation}
%====================================================
with $\tilde T_e=k_B T_e/ \hbar\omega_c$ and 
%====================================================
\begin{equation}
A_L ={4 \sqrt{2} \over \pi^{5/2} h } {\lvh \over l^3} (D \beta)^2 \vt^3 I_p
\end{equation}
%====================================================
with $\beta=(u_0/l)(\omega_c / \omega_0)$, 
$u_0^2=\hbar /\rho_m c_l l^2$, and
$\omega_0=c_l/ l$. 
$I_p$ is defined by 
%====================================================
\begin{equation}
I_p =\int^{\infty}_0 dx {x^2 \over (1+a^2 x^2 )^3} \exp(-x^2/2)
\end{equation}
%====================================================
with $a=\vt \omega_c /c_l b$. 

Phonon scatterings (absorptions and emissions) also give 
an electron hopping between a hill and a valley 
and contribute to the energy gain. 
However, the energy gain due to phonon scatterings is shown to be 
$10^{-3}$ of that due to Coulomb scatterings 
even at $T_L=T_e$. 
%#############################################################
\begin{figure}
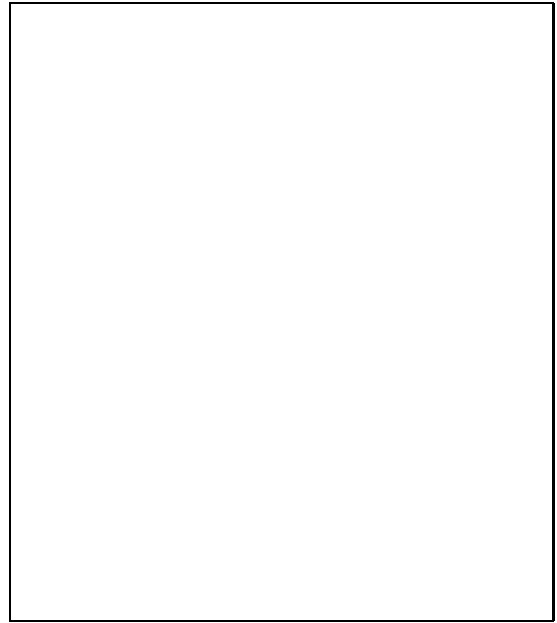

\figureheight{8cm}
\caption{
(a) The energy gain $P_G$ at three values of the electric field $E$ 
and the loss $P_L$ (both divided by $A_L I_f$) 
as a function of the electron temperature $T_e$ 
($\tilde T_e=k_B T_e/ \hbar\omega_c$). 
Points of intersection indicates steady states. 
(b) $T_e$ and $\sigma_{xx}$
($\sigma_{xx}/\sigma_0 I_f= 4f(0)[1-f(0)]$ with $\sigma_0$ the constant) 
in steady states as a function of $E$
(both have the lower branch of $T_e=0$ and $\sigma_{xx}=0$). 
$P_G(E)=P_L(E) \propto T_e(E)$ 
if the weak $T_e$ dependence of $I_f$ is neglected. 
} 
\label{fig:bistability}
\end{figure}
%#############################################################

The energy gain $P_G$ and the loss $P_L$ are 
plotted as a function of the electron temperature $T_e$ 
in Fig.~\ref{fig:bistability}(a). 
The points of intersection ($P_G=P_L$) give $T_e$ in steady states. 
The number of points of intersection increases from one to three 
as increasing the electric field $E$ through a critical value $E_{c1}$, 
and a bistability appears above $E_{c1}$ 
(one of three in the middle corresponds to an unstable state). 
The nonlinear $T_e$ dependence of $P_G$ giving this bistability 
is due to the presence of the activation energy for 
the current-carrying states at the Landau-level center. 
$T_e=0$ is a stable solution at any $E$ 
and the upper critical field is $E_{c2}=\infty$,  
since $P_G=P_L=0$ at $T_e=0$
in the present approximations of $T_L=0$ and 
no tunneling conduction. 
$T_e$ in steady states as a function of $E$ is given by 
the equation $P_G=P_L$ which is simplified to
%====================================================
\begin{equation}
{A_G \over A_L} \tilde E^2 = g(\tilde T_e) \equiv  
\tilde T_e [ \exp(1 / 2\tilde T_e ) + \exp(-1 / 2\tilde T_e ) +2 ] , 
\end{equation}
%====================================================
and is plotted with $\sigma_{xx}(E)=P_G/E^2$ in Fig.~\ref{fig:bistability}(b). 
At the critical point, 
$\tilde T_{ec} \equiv \tilde T_{e}(E_{c1})= 0.32$, 
and $A_G \tilde E_{c1}^2 / A_L =2.2$. 

The estimate of the critical electron temperature is 
free from ambiguities in values of $A_G$ and $A_L$ 
and is given by $T_{ec}=0.32\hbar\omega_c/k_B$. 
Unfortunately, there exists no direct experimental estimate of $T_e$. 
Indirect estimates have been made from the energy balance equation 
using the observed temperature dependence of $\sigma_{xx}$ at $E \ll E_{c1}$
and a calculated energy loss.~\cite{Komiyama85,Scherer97}
Theoretical and experimental estimates of $T_{e}$ at $E_{c1}$ are 
%====================================================
\begin{eqnarray}
T_{ec}({\rm theory}) &=& 32{\rm K}
\ \ (B=5{\rm T}), \nonumber \\
T_{ec}({\rm ref.}\citen{Komiyama85}) &=& 8{\rm K} 
\ \ (B=3.8{\rm T}), \nonumber \\
T_{ec}({\rm ref.}\citen{Scherer97}) &=& 15{\rm K} 
\ \ (B=7.6{\rm T}). \nonumber 
\end{eqnarray}
%====================================================
The discrepancy between the theory and the experiments is reduced 
if we consider in the theory 
the nonzero lattice temperature and 
the smaller activation energy $E_{\rm ac}$. 
In the present theory $E_{\rm ac}=\hbar \omega_c /2$, 
while $E_{\rm ac}=0.7 \hbar \omega_c /2$ 
from the observed $\sigma_{xx}(T)$.~\cite{Komiyama85}

The estimate of the lower critical field $E_{c1}$ depends on $A_L/A_G$. 
We use $B=5{\rm T}$, $\lvh/l=10$, $b=0.03 {\rm \AA}^{-1}$, $\vt=0.1$, 
$D$=10eV, $\rho_m=5.3 \times 10^3{\rm kg/m}^3$, 
$c_l=5\times 10^3{\rm m/s}$,  $N_c=2$, $f_L=0.1$ and 
$<I_{\rm orb}^2>^{1/2}=0.4$, 
and assume that $\rho(\ve)=$const. Then we obtain 
%====================================================
\begin{eqnarray}
E_{c1}({\rm theory})&=&20{\rm V/cm} 
\ \ (B=5{\rm T}), \nonumber \\
E_{c1}({\rm ref.}\citen{Ebert83})&=&65{\rm V/cm} 
\ \ (B=4.7{\rm T}), \nonumber \\
E_{c1}({\rm ref.}\citen{Komiyama85}) &=&40{\rm V/cm} 
\ \ (B=3.8{\rm T}), \nonumber \\
E_{c1}({\rm ref.}\citen{Scherer97}) &=&100{\rm V/cm} 
\ \ (B=7.6{\rm T}). \nonumber 
\end{eqnarray}
%====================================================
The order of magnitude agrees between the theory and the experiments. 
Since there are large ambiguities in values of 
$ \lvh/l$, $N_c$, and $f_L$, 
the discripancy between the theory and the experiments 
is within the limitation of accuracy
($E_{c1} \propto N_c^{-1/2}$ and $N_c$ takes a smaller value 
in a sparse network of conduction channels between hills and valleys, 
which is probable in disordered potentials~\cite{Nixon90}).
Note that $E_{c1}$ in the real system is larger than
the present theoretical estimate, 
since we have neglected fluctuations around the homogeneous steady state. 

The estimate of the energy dissipation at $E_{c1}$ is given by 
$P_G(T_{ec}, E_{c1})=P_L(T_{ec})$, 
and depends on $A_L$ and $I_f$. 
Theoretical and experimental estimates of $P_G$ at $E_{c1}$ are 
%====================================================
\begin{eqnarray}
P_{Gc}({\rm theory})&=&
5\times 10^{-2}{\rm J s^{-1} cm^{-2}} 
\ \ (B=5{\rm T}), \nonumber \\
P_{Gc}({\rm ref.}\citen{Ebert83})&=&
1.5\times 10^{-2}{\rm J s^{-1} cm^{-2}} 
\ \ (B=4.7{\rm T}), \nonumber \\
P_{Gc}({\rm ref.}\citen{Komiyama85}) &=&
0.8\times 10^{-2}{\rm J s^{-1} cm^{-2}} 
\ \ (B=3.8{\rm T}), \nonumber \\
P_{Gc}({\rm ref.}\citen{Scherer97}) &=&
0.6\times 10^{-2}{\rm J s^{-1} cm^{-2}}
\ \ (B=7.6{\rm T}). \nonumber 
\end{eqnarray}
%====================================================
The agreement in $P_{Gc}$ is poorer than in $E_{c1}$,  
possibly because $P_{Gc}$ depends stronger on $A_L$. 

In conclusion, 
we have considered the Coulomb scattering between 
localized orbits as an electronic process for the energy dissipation 
in the breakdown of the quantum Hall effect 
in the presence of slowly-fluctuating potentials.
Compared to the previous theories 
based on inter-Landau-level phonon scatterings, 
we have obtained better agreements with the experiment 
in the value of the lower critical electric field, and 
in the existence of the bistability.

%++++++++++++++++++++++++++++++++++++++++++++++++++++
The author would like to thank 
T.\ Ando, S.\ Komiyama, N.\ Tokuda, and Y.\ Asano for valuable discussions. 

%++++++++++++++++++++++++++++++++++++++++++++++++++++

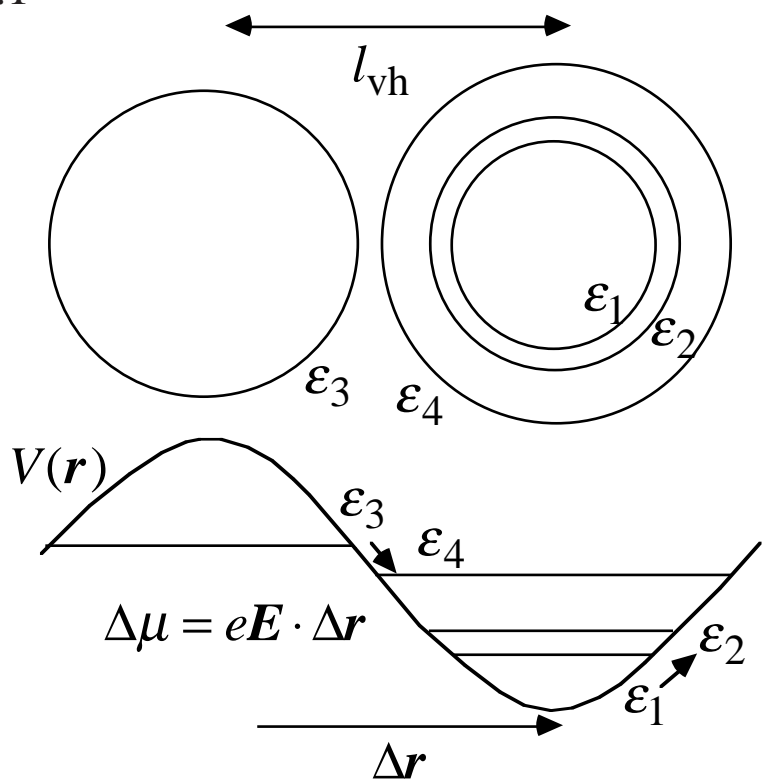
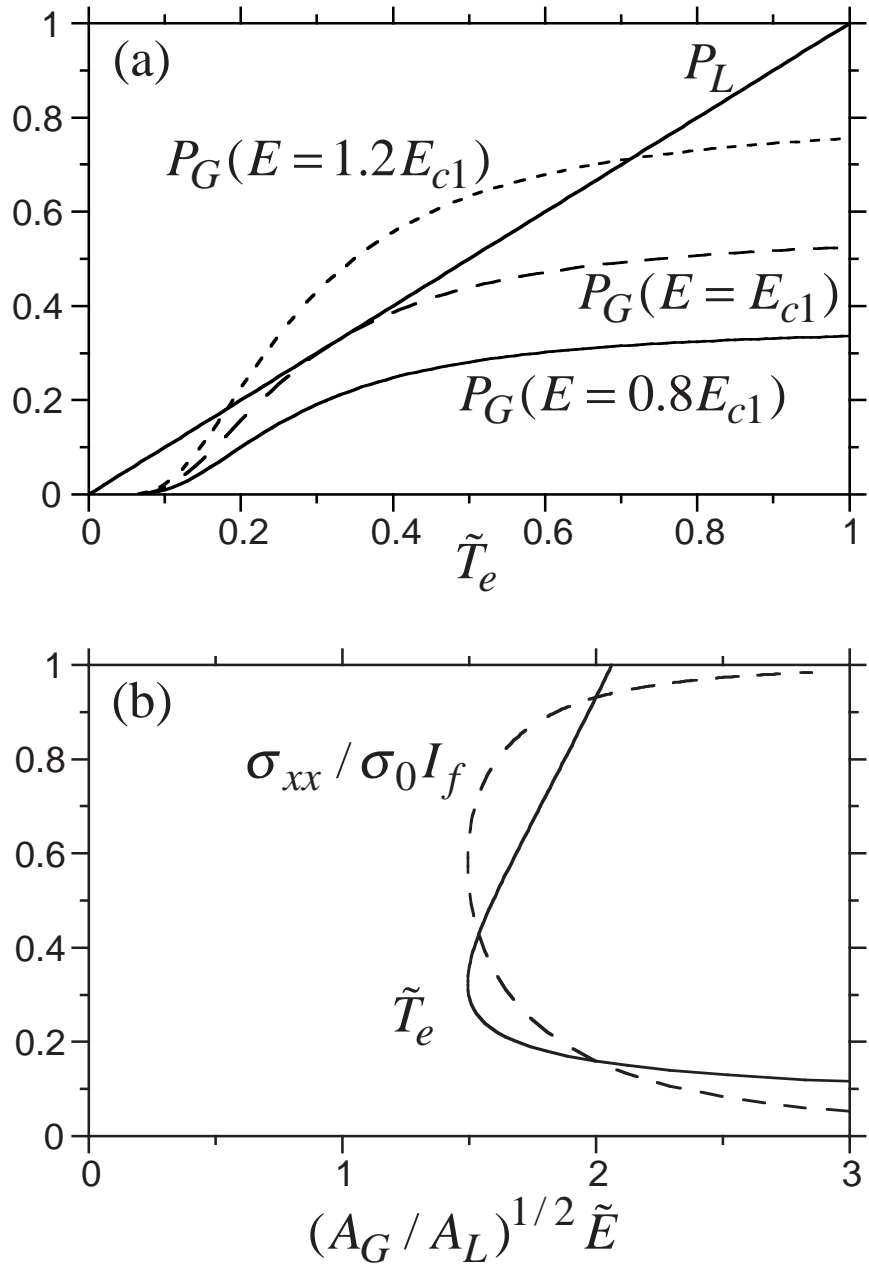
\end{document}